\documentclass[twocolumn,aps,amsmath,amssymb,showpacs,showkeys,floatfix,prl,superscriptaddress]{revtex4}

\usepackage{bbm}
\usepackage[dvips]{graphicx}
\usepackage{color}
\usepackage{amsmath,amssymb,latexsym}

\newcommand{\micron}{\ensuremath{\mu\mathrm{m}}}
\newcommand{\won}{\omega_\mathrm{on}}
\newcommand{\woff}{\omega_\mathrm{off}}
\renewcommand{\r}{\mathbf{r}}
\newcommand{\R}{\mathbf{R}}
\newcommand{\e}{\mathbf{e}}
\newcommand{\eps}{\varepsilon}
\renewcommand{\O}{\mathcal{O}}
\renewcommand{\Re}{\mathrm{Re}\,}
\renewcommand{\ol}[1]{\overline{#1}}

\newcommand{\lsim}{\mathrel{\hbox{\rlap{\lower.55ex\hbox{$\sim$}} \kern-.3em \raise.4ex \hbox{$<$}}}}
\newcommand{\gsim}{\mathrel{\hbox{\rlap{\lower.55ex\hbox{$\sim$}} \kern-.3em \raise.4ex \hbox{$>$}}}}
\newcommand{\dA}{\delta\! A}

\def \omegaoff {\omega_{\mathrm{off}}}
\def \omegaon {\omega_{\mathrm{on}}}

\newcommand{\rev}[1]{{#1}} 

\begin{document}

\title{Active phase and amplitude fluctuations of flagellar beating\\
(accepted for publication in Physical Review Letters)}

\author{Rui Ma}
\affiliation{Max Planck Institute for the Physics of Complex Systems, Dresden, Germany}
\affiliation{Institute for Advanced Study, Tsinghua University, Beijing, China}
\author{Gary S. Klindt}
\affiliation{Max Planck Institute for the Physics of Complex Systems, Dresden, Germany}
\author{Ingmar H. Riedel-Kruse}
\affiliation{Department of Bioengineering, Stanford University, Stanford, CA, USA}
\author{Frank J{\"u}licher}
\affiliation{Max Planck Institute for the Physics of Complex Systems, Dresden, Germany}
\author{Benjamin M. Friedrich}
\affiliation{Max Planck Institute for the Physics of Complex Systems, Dresden, Germany}
\email{benjamin.friedrich@pks.mpg.de}

\date{\today}

\bibliographystyle{apsrev}



\begin{abstract}
The eukaryotic flagellum beats periodically, 
driven by the oscillatory dynamics of molecular motors,
to propel cells and pump fluids.
Small, but perceivable fluctuations in the beat of individual flagella
have physiological implications for synchronization in collections of flagella
as well as for hydrodynamic interactions between flagellated swimmers.
Here, 
we characterize phase and amplitude fluctuations of flagellar bending waves
using shape mode analysis and limit-cycle reconstruction.
We report a quality factor of flagellar oscillations, $Q=38.0\pm 16.7$ (mean$\pm$s.e.). 
Our analysis shows that flagellar fluctuations are dominantly of active origin.
Using a minimal model of collective motor oscillations,
we demonstrate how the stochastic dynamics of individual motors
can give rise to active small-number fluctuations in motor-cytoskeleton systems.
\end{abstract}

\maketitle

Systems far from equilibrium such as living matter 
display active, non-thermal fluctuations, 
as well as directed motion and oscillations, which are important for biological function.
As a prominent example, 
molecular motors coupled to cytoskeletal filaments
convert chemical energy into work and heat
to generate motion at the cellular scale.
Motor-filament systems can drive mechanical oscillations
including 
spontaneous hair bundles oscillations in the ear \cite{Martin:2003},
mitotic spindle oscillations during cell division \cite{Grill:2005}, 
sarcomere oscillations in insect flight muscle \cite{Jewell:1966},
and the regular bending waves of cilia and flagella,
which propel cells in a liquid including sperm and green algae \cite{Gray:1928}, 
as well as clear mucus in mammalian airways \cite{Sanderson:1981}.
Cilia and flagella are slender cell appendages of $10{-}100\,\micron$ length,
ubiquitously found in non-bacterial cells, 
which comprise a conversed cylindrical scaffold of microtubules
interspersed by dynein molecular motors.

The collective dynamics of the motors 
working against a visco-elastic load
drives flagellar oscillations via a dynamic instability \cite{Julicher:1997}.
Force generation by individual motors relies on the stochastic progression through a mechanico-chemical cycle \cite{Howard:2001}.
The stochastic nature of force generation should manifest itself
in oscillations that display a characteristic level of noise, representative of active fluctuations.
Intriguingly, previous work reported 
Fourier peaks of finite width in power spectra of flagellar oscillations \cite{Riedel:2005a}, and
phase-slips in pairs of synchronized flagella \cite{Polin:2009,Goldstein:2009,Wan:2013},
which allowed an indirect assessment of flagellar noise.
A direct measurement of flagellar fluctuations is pending, let alone a mechanistic understanding.
Flagellar fluctuations impart on biological function:
Phase fluctuations of flagellar beating should counter-act synchronization in collections of flagella,
which is important for fast swimming \cite{Brennen:1977}
and efficient fluid pumping \cite{Osterman:2011}.
Amplitude fluctuations will result in noisy swimming paths of flagellated swimmers
and impart on hydrodynamic interactions between swimmers
\cite{Drescher:2011}. 

Here, we report direct measurements of phase and amplitude fluctuations of the flagellar beat
and discuss the microscopic origin of active flagellar fluctuations using a minimal model.
We further illustrate the impact of flagellar fluctuations on swimming and synchronization.
Our analysis contributes to a recent interest in 
driven, out-of-equilibrium systems and their fluctuation fingerprint 
\cite{Betz:2009,Placais:2009,Isaac:2011,Otten:2012} 
by characterizing noisy limit-cycle dynamics in an ubiquitous motility system, the flagellum.

\begin{figure}
\begin{center}
\includegraphics[width=8.5cm]{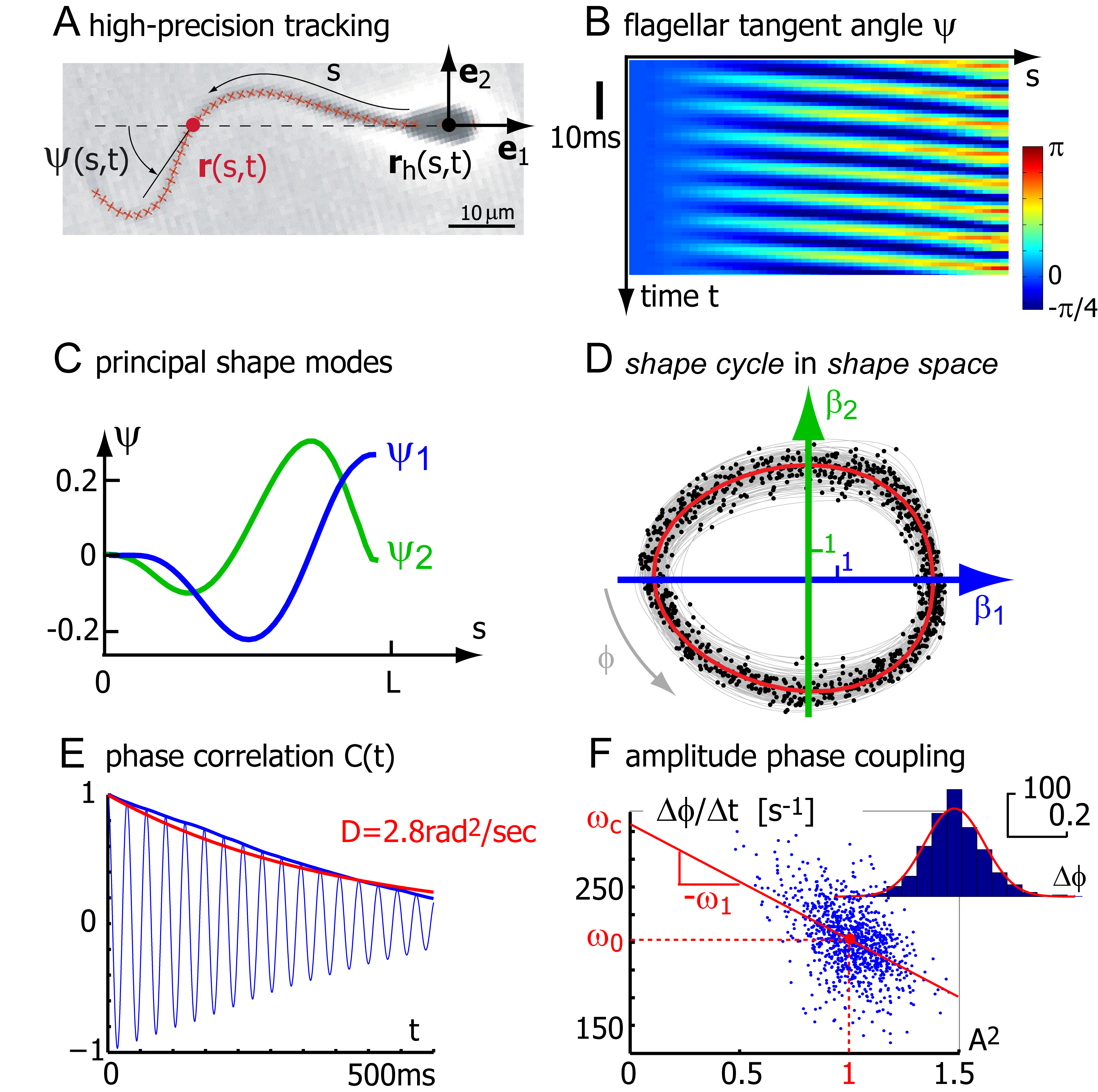}
\end{center}
\caption[]{
\label{fig_sperm}
(color online) 
The flagellar beat of sperm cells displays active fluctuations.
\textbf{A}. 
Tracked flagellar shapes are conveniently characterized by a tangent angle $\psi(s,t)$.
\textbf{B}. 
The kymograph of this tangent angle reveals the periodicity of the flagellar beat.
\textbf{C}. 
Using principal component analysis, 
we identify two principal shape-modes $\psi_1(s)$, $\psi_2(s)$,
whose superpositions account for 95\% of the variability of the tangent angle data.
\textbf{D}. 
By projecting the tangent angle data on the shape-space spanned by $\psi_1(s)$ and $\psi_2(s)$,
each flagellar shape is assigned a pair of shape-coefficients, $(\beta_1,\beta_2)$, see eq.~(\ref{eq_beta}).
This representation allows us to define a limit-cycle of perfect periodic beating (red).
By projection onto this limit-cycle, 
we define a phase $\varphi$ for each flagellar shape.
\textbf{E}. 
The flagellar phase-diffusion coefficient $D$ is determined
by fitting an exponential decay (red) to 
the phase correlation function 
($|C(t)|$: blue, $\Re C(t)$: light blue).
\textbf{F}.
Phase speed $\Delta\varphi_i/\Delta t$ and squared amplitude $A(t_i)$ are negatively correlated.
Inset: phase increments are approximately normally distributed.
}
\end{figure}

\textit{Flagellar shape analysis.}
We characterize flagellar beat patterns as superposition of principal shape modes.
This dimensionality reduction is key to our fluctuation analysis.
We analyze planar beat patterns of bull sperm
swimming close to a boundary surface \cite{Riedel:2007},
filmed at $250$ frames-per-second
\rev{(corresponding to about $8$ frames per beat cycle)}.
The flagellar centerline $\r(s,t)$,
tracked as function of arclength position $s$ and time $t$,
can be expressed with respect to a material frame of the sperm head
in terms of a tangent angle $\psi(s,t)$
\begin{equation}
\r(s,t) = \r_h(t) - \int_0^s ds'\, [ \cos\psi(s',t)\e_1+\sin\psi(s',t)\e_2 ].
\end{equation}
Here, $\r_h(t)$ denotes the sperm head center, 
and $\e_1$ and $\e_2$ are ortho-normal vectors 
with $\e_1$ pointing along the long head axis, see Fig.~\ref{fig_sperm}A.
A space-time-plot of $\psi(s,t)$ reveals the periodicity of the flagellar beat, see Fig.~\ref{fig_sperm}B.
This high-dimensional data set can be projected on a low dimensional `shape space'
using shape mode analysis based on principal component analysis \cite{Geyer:2013}. 
The time-averaged tangent angle $\psi_0(s){=}\sum_{i=1}^n \psi(s,t_i)/n$
characterizes the mean shape of the beating flagellum ($n{=}1024$ frames in each movie).
We further define a two-point correlation matrix
$M(s,s'){=}\sum_i [\psi(s,t_i){-}\psi_0(s)][\psi(s',t_i){-}\psi_0(s')]$,
where $s$, $s'$ range over $m$ equidistant arc-length positions along the flagellum.
The eigenvectors $\psi_j(s)$ of the symmetric $m\times m$-matrix $M$, 
sorted by decreasing magnitude of the corresponding eigenvalues, 
characterize principal shape modes of the flagellar beat.
The first two shape modes account for $95{\pm}1\%$ of the variance of the tangent angle data
(all measurements are mean$\pm$s.e., $n{=}7$ cells).
We project the full data set on 
a two-dimensional shape-space spanned by these two shape-modes
\begin{equation}
\label{eq_beta}
\psi(s,t)\approx \psi_0(s) + \beta_1(t)\psi_1(s)+\beta_2(t)\psi_2(s)
\end{equation}
with shape-coefficients $\beta_1$, $\beta_2$ obtained by least-square fit, see Fig.~\ref{fig_sperm}C,D.
Flagellar beating implies periodic shape changes of the flagellum,
and thus noisy oscillations of the shape-coefficients
with mean frequency $\omega_0{=}2\pi/T$,
where $T{=}32.4{\pm}1.9\,\mathrm{ms}$.
Individually, $\beta_1(t)\psi_1(s)$ and $\beta_2(t)\psi_2(s)$ describe standing waves;
their combination results in a traveling wave propagating from the base to the tip of the flagellum, 
thereby facilitating net propulsion.

\textit{Limit-cycle reconstruction.}
The point cloud representing subsequent flagellar shapes in Fig.~\ref{fig_sperm}D forms a closed loop.
This allows us to define a limit-cycle of noisy flagellar oscillations (red)
by fitting a closed curve $\left(\ol{\beta}_1(\varphi),\ol{\beta}_2(\varphi)\right)$,
parametrized by a phase $\varphi$.
The phase parametrization of the limit-cycle is defined such that 
the mean of the phase speed is independent of $\varphi$ \cite{Kralemann:2008}.
Thus, $\varphi$ slightly differs from the polar angle in the $(\beta_1,\beta_2)$-plane.
Next, we assign a unique flagellar phase to each tracked flagellar shape 
by projecting the corresponding point in the $(\beta_1,\beta_2)$-plane
radially onto the limit-cycle.
The shape trajectory $\left(\beta_1(t),\beta_2(t)\right)$ 
avoids the singular origin,
thus the instantaneous phase speed $\dot{\varphi}$ is well-defined.

\textit{Phase fluctuations.}
The phase speed $\dot{\varphi}$ has mean equal to the frequency $\omega_0$ of the beat,
but can fluctuate around this mean.
Phase speed fluctuations 
cause a decay of the phase-correlation-function
$C(t){=}\langle \exp[i[\varphi(t_0+t){-}\varphi(t_0)]\rangle]$,
see Fig.~\ref{fig_sperm}E.
This decay is insensitive to measurement noise that is uncorrelated from frame to frame.
The frame-to-frame phase increments $\Delta\varphi_i{=}\varphi(t_{i+1}){-}\varphi(t_i)$
are approximately normally distributed (Fig.~\ref{fig_sperm}F, inset).
Further, the correlation time of phase speed fluctuations 
is on the order of our temporal resolution $4\,\mathrm{ms}$ or below,
and thus short compared to the time-scale of phase decoherence.
We can thus interpret the observed phase decoherence 
using an idealized model of $\delta$-correlated phase speed fluctuations,
\begin{equation}
\label{eq_phi}
\dot{\varphi}=\omega_0+\zeta,
\end{equation}
where $\zeta$ is Gaussian white noise with $\langle\zeta(t)\zeta(t')\rangle{=}2D\delta(t-t')$
and $D$ denotes a phase-diffusion coefficient.
In this idealization, $|C(t)|{=}\exp({-}D|t|)$.
By fitting an exponential to measured $|C(t)|$,
we obtain the phase-diffusion coefficient of sperm flagellar beating,
$D{=}3.2{\pm}1.9\,\mathrm{s}^{-1}$, see Fig~\ref{fig_sperm}E.
An alternative measure for the phase stability of oscillations is the quality factor,
$Q=\omega_0/(2D)=38.0{\pm}16.7$,
where $\omega_0/Q$ indicates 
the width at half-maximum of the principal peak in the power spectral density \rev{of ${\exp}[i\varphi(t)]$}.

The observed phase fluctuations of the flagellar beat are dominantly of active origin
and surpass passive, thermal fluctuations by orders-of-magnitude 
(as suggested by earlier, indirect measurements \cite{Goldstein:2009}):
For a simple estimate,
we consider a flagellar beat that is constrained to move along the shape limit-cycle
with $\varphi$ as only degree of freedom.
The friction force $P_\varphi$ conjugate to $\varphi$ 
comprises hydrodynamic friction $\gamma\dot{\varphi}$ and dissipation within the flagellum.
We estimate $\gamma{\approx} 3\,\mathrm{pN}\,\micron\,\mathrm{s}$ \cite{Gray:1955b,Friedrich:2010}.
We thus obtain an upper bound
$k_B T/\gamma \approx 0.0015\,\mathrm{s}^{-1}$
for the contribution of thermal fluctuations to phase-diffusion $D$, 
which is a thousand-fold smaller than the value measured.

\textit{Amplitude fluctuations.}
We define an instantaneous amplitude of the flagellar beat,
$A(t){=}|\beta_1(t)+i\beta_2(t)|/\rho_0\left(\varphi(t)\right)$,
normalized by $\rho_0(\varphi){=}|\ol{\beta}_1(\varphi)+i\ol{\beta}_2(\varphi)|$.
Thus, the complex oscillator variable
$Z(t){=}A(t)e^{i\varphi(t)}$
maps the shape limit-cycle onto the unit circle.
In our data, the amplitude $A(t)$ is approximately normally distributed
with $\sigma_A^2 = \langle A(t)^2\rangle{-}1{=}0.0070{\pm}0.0023$
\endnote{
The contribution from measurement noise is small.
As a test, we added random perturbations to $\r(s,t)$, using known accuracies of tracking \cite{Riedel:2007}.
Phases and amplitudes for perturbed and unperturbed data were strongly correlated;
results for $\sigma_A$ did not significantly change.
}.
The autocorrelation function of amplitude fluctuations
decays with time-constant $\tau_A{=}5.9{\pm}1.8\,\mathrm{ms}$.
Interestingly, 
we find that phase speed correlates with amplitude squared;
the slope $-\omega_1$ of a linear regression gives 
$\omega_1/\omega_0{=}0.38{\pm}0.10$,
see Fig.~\ref{fig_sperm}F.
Thus, the beating flagellum is represented as a non-isochronous oscillator 
(with approximate isochrones $\varphi{-}2\tau_A\omega_1{\rm ln}A{=}{\rm const}$ \cite{Pikovsky:synchronization}).
Non-isochrony of non-linear oscillators has been related to synchronization 
\rev{\cite{Niedermayer:2008,Leoni:2012}}.

\textit{Noisy normal form.}
Previous theoretical work described the onset of flagellar oscillations
as a supercritical Hopf bifurcation \cite{Camalet:2000}
with normal form ($\mu{>}0$) \cite{Crawford:1991}
\begin{equation}
\label{eq_Hopf}
\dot{Z}=
i(\omega_c-\omega_1|Z|^2)Z+\mu(\Lambda-|Z|^2)Z+\Xi.
\end{equation}
In the absence of noise, $\Xi{=}0$, as considered originally \cite{Camalet:2000},
the complex oscillator variable
$Z(t){=}A(t)e^{i\varphi(t)}$ exhibits spontaneous oscillations
with amplitude $A{=}\Lambda^{1/2}$ 
and frequency $\omega_0{=}\omega_c{-}\omega_1\Lambda$
for effective motor activity $\Lambda{>}0$.
In this case, we may assume $\Lambda{=}1$ after a parameter transformation.

To study the role of fluctuations, 
we add a multiplicative noise term
$\Xi{=}Z(\zeta_A+i\zeta_\varphi)$ with
uncorrelated Gaussian white noise variables satisfying
$\langle\zeta_j(t)\zeta_k(t')\rangle=2D_{j}\delta_{jk}\delta(t-t')$, $j,k\in\{A,\varphi\}$,
and use Stratonovich interpretation.
This choice represents the simplest phase-invariant noise term 
with tunable phase and amplitude noise strengths $D_\varphi$ and $D_A$ \cite{Graham:1982}.
For weak noise, $D_A,D_\varphi{\ll}\mu\Lambda$,
amplitude fluctuations satisfy
$\langle A(t_0)A(t_0+t)\rangle-1\approx\sigma_A^2\exp(-|t|/\tau_A)$
with correlation-time $\tau_A{=}(2\mu\Lambda)^{-1}$ and variance $\sigma_A^2{=}D_A\tau_A\Lambda$.
Phase fluctuations are colored with effective phase-diffusion coefficient
$D{=}D_\varphi{+}(\omega_1/\mu)^2 D_A$.
Our measurements of active flagellar fluctuations thus allow the full parametrization of eq.~(\ref{eq_Hopf})
(with $\Lambda{=}1$).
Note that in the special case $D_A{=}D_\varphi\ll\mu\Lambda$,
our choice of multiplicative noise gives the same long-term behavior as additive noise.

\textit{Flagellar fluctuations imply non-deterministic swimming:}
Using measured noise strengths, 
we simulated realistic beat patterns and corresponding stochastic swimming paths, see Fig.~\ref{fig_path}A.
Specifically, we 
(i) use eq.~(\ref{eq_Hopf}) to simulate $Z(t){=}A(t)e^{i\varphi(t)}$,
(ii) construct shape coefficients 
$\beta_1(t)+i\beta_2(t)=A(t)\rho_0(\varphi(t))$, 
and tangent angles $\psi(s,t)$ by eq.~(\ref{eq_beta}),
(iii) compute the path $\r_h(t)$ using resistive force theory \cite{Gray:1955b} 
as described in \cite{Friedrich:2010}.
We find that the center $\R(t)$ of sperm swimming circles diffuses with diffusion coefficient
$D_\mathrm{R}{=}3.3\,\micron^2/\mathrm{s}$, which is on the same order of magnitude, albeit smaller,
than a value $D_\mathrm{R}{=}9{\pm}2\,\micron^2/\mathrm{s}$ 
measured for sea urchin sperm \cite{Riedel:2005a}.
Our analysis includes amplitude and phase fluctuations, but neglects additional shape fluctuations;
thus our value is a lower bound.

Although phase and amplitude fluctuations are correlated,
we can ask separately for their respective effect on swimming.
Phase fluctuations cause fluctuations in swimming speed, but do not change the shape of the path.
This is because the Stokes equation governing self-propulsion at low Reynolds numbers \cite{Lauga:2009}
is invariant under (stochastic) re-parametrizations of time.

To gain qualitative insight into the microscopic origin of noisy oscillations, 
and the dependence of phase-diffusion on microscopic parameters,
we now discuss a minimal motor model and show how it can be mapped onto eq.~(\ref{eq_Hopf}).

\begin{figure}
\begin{center}
\includegraphics[width=8.5cm]{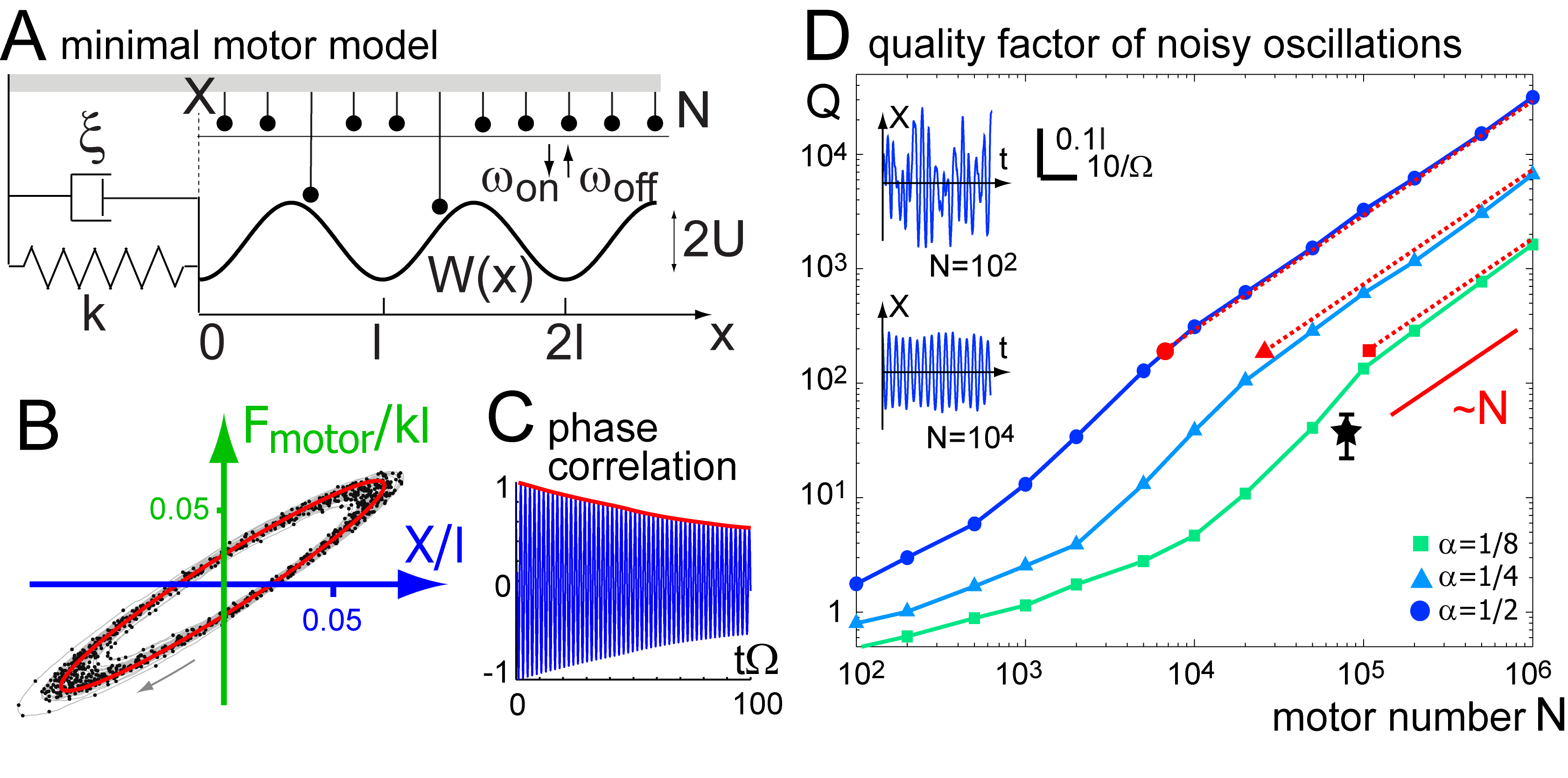}
\end{center}
\caption[]{
\label{fig_motor_model}
(color online) 
A minimal model of coupled motors predicts noisy oscillations.
\textbf{A.} 
An ensemble of $N$ motors, grafted at a rigid backbone (gray),
can bind and unbind to a filament with transition rates $\won$ and $\woff$.
Bound motors interact with the filament through an interaction potential $W(x)$.
Filament and backbone are coupled visco-elastically.
\textbf{B.}
The motor model exhibits spontaneous, noisy oscillations,
here visualized 
by filament position $X$ and total motor force.
The deterministic limit-cycle is shown in red.
\textbf{C.}
The phase correlation function $C(t)$
(real-part shown in blue)
decays exponentially,
$|C(t)|{\approx}{\rm exp}(-Dt)$,
defining the phase-diffusion coefficient $D$.
\textbf{D.}
The quality factor $Q{=}\omega_0/(2D)$ scales with $N$ for large $N$,
consistent with our analytic approximation (dashed red, eq.~(\ref{eq_D})).
The star indicates the experimentally measured $Q$.
For all simulations, we chose parameters close to the Hopf bifurcation
($\xi_a/\xi{=}1.2\pi^2$, $\nu{=}10$, $\alpha{=}\eta{=}0.5$, $N{=}10^4$,
unless indicated otherwise; errors smaller than symbol size).
}
\end{figure}

\begin{figure}
\begin{center}
\includegraphics[width=8.5cm]{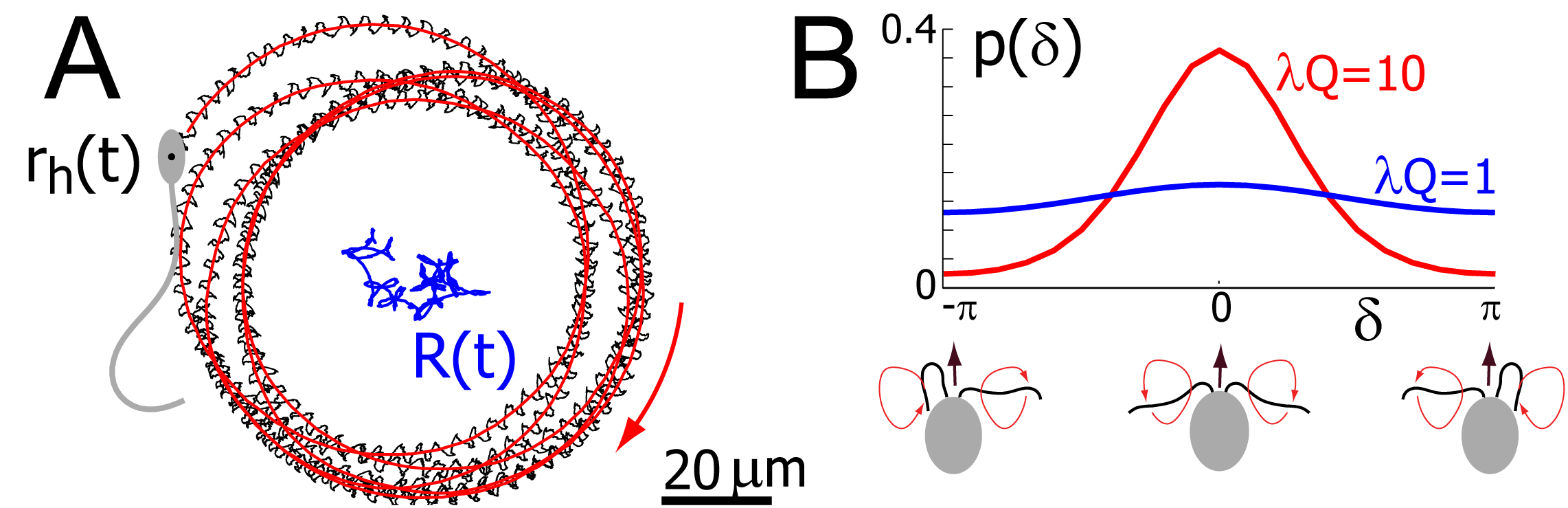}
\end{center}
\caption[]{
\label{fig_path}
(color online)
Flagellar fluctuations imply
non-deterministic swimming and counter-act synchronization.
\textbf{A}.
We simulated stochastic sperm swimming paths $r_h(t)$ (black),
using measured flagellar fluctuation strengths.
Fluctuations imply that the blue center $\mathrm{R}(t)$ of red sperm swimming circles diffuses,
with apparent diffusion coefficient $D_\mathrm{R}{=}3.30{\pm}0.01\,\micron^2/\mathrm{s}$.
\textbf{B}.
Pairs of flagella can synchronize, \textit{e.g.} in the green alga \textit{Chlamydomonas}.
In a simple description of flagellar synchronization,
the phase difference $\delta$ between its two flagella peaks around zero
for realistic noise strength ($\lambda Q{=}10$),
but is almost uniformly distributed for ten-fold stronger noise ($\lambda Q{=}1$),
indicating lack of synchronization.
}
\end{figure}

\textit{A minimal model for noisy motor oscillations.}
We exemplify how a finite collection of motors
drives spontaneous oscillations with characteristic small-number-fluctuations 
using the classical two-state model \cite{Julicher:1997,Guerin:2011a} in its most simple form:
A collection of $N$ motors, rigidly attached to an inextensible backbone
interacts with a filament through an effective potential,
$W(x){=}U[1-\cos(2\pi x/l)]$,
see Fig.~\ref{fig_motor_model}A.
Here $x$ is the coordinate of the motor along the filament,
and $l$ the periodicity of the filament.
Individual motors can bind to and unbind from the filament
with rates
$\won(x){=}\Omega[\eta{-}\alpha\cos(2\pi x/l)]$
and $\woff{=}\Omega{-}\won$.
Here, $\eta$ denotes the mean fraction of attached motors (``duty ratio'').
Importantly, the binding rates are spatially inhomogeneous,
characterized by $\alpha$, and break detailed balance.
If the filament is now coupled to the backbone
by a visco-elastic element with viscosity $\xi$ and elastic stiffness $k$,
we obtain a force-balance equation for the position $X(t)$ of the filament,
$kX{+}\xi\dot{X}{=}F_\mathrm{m}$ with $F_\mathrm{m}={-}\sum_{i} \partial_X W(x_i{-}X)$,
where the sum extends over all bound motors and 
$x_i{=}il/N$ is a simple choice for the positions of the motors along the backbone.

To properly define a thermodynamic limit for large $N$,
we rescale stiffness and viscosity as $k{=}k_0 N$ and $\xi{=}\xi_0 N$.
In the limit $N{\rightarrow}\infty$, 
the system can exhibit spontaneous oscillations by a supercritical Hopf-bifurcation, 
when the normalized motor activity $\xi_a/\xi{=}2\pi^2\alpha N U/(\Omega l^2\xi)$
exceeds the threshold $1{+}\nu$, where $\nu{=}k/(\xi\Omega)$ \cite{Guerin:2011a}.
For a finite motor number, we numerically observe noisy oscillations, see Fig.~\ref{fig_motor_model}.

In the following, we analytically compute the quality factor $Q$
in the limit of large (yet finite) motor number $N$, 
assuming that we are close to the Hopf bifurcation with $\eps=\xi_a/\xi{-}1{-}\nu$ positive and small.
Following \cite{Guerin:2011a,Guerin:2011b}, 
we first approximated the stochastic binding and unbinding dynamics of individual motors
by a diffusion approximation, thus arriving at a Fokker-Planck equation for the 
probability distribution of filament position and density $\rho(x)$ of bound motors,
see appendix for details.
Due to the simple choice of potential $W(x)$, 
the dynamics of the first Fourier mode of $\rho(x)$ decouples from that of the higher modes,
resulting in a 3-dimensional stochastic system \cite{Guerin:2011b}.
A non-linear coordinate transformation maps this system 
onto Hopf normal form eq.~(\ref{eq_Hopf}),
with oscillator variable $Z$ satisfying $\Re Z=X/l{+}\O(\eps^{3/2})$
and phase-dependent noise term $\Xi{=}i\zeta$, 
where 
$\langle\zeta(t)\zeta(t')\rangle=4D\Lambda\,\delta(t-t')$.
The quality factor $Q=\omega_0/(2D)$ is found to scale with $N$
\begin{equation}
\label{eq_D}
Q\approx \frac{\omega_0}{2\Omega}\frac{N\Lambda}{\eta(1-\eta)}
\left( \frac{2\pi\alpha}{\sqrt{\nu}+1/\sqrt{\nu}} \right)^2.
\end{equation}
Furthermore,
$\Lambda{\approx}\eps(1+4\nu)/[3\pi^2\nu(1+2\nu)]$, 
$\mu{\approx}\Omega\eps/(2\Lambda)$,
$\omega_0{\approx}\Omega\sqrt{\nu}[1+\eps/(2+4\nu)]$.
Interestingly, 
the motor duty ratio $\eta$ controls oscillation quality, 
although $\eta$ affects neither amplitude nor frequency (for $N{\rightarrow}\infty$).
To understand this, note that the number of bound motors 
fluctuates with mean $\eta N$ and variance $\eta(1{-}\eta)N$.
This number characterizes a spatially homogeneous ``background'' of bound motors, 
which does not contribute directly to the oscillations,
but sets the amplitude of motor density fluctuations responsible for phase-diffusion.
Oscillations become also more regular for increasing amplitude.
Eq.~(\ref{eq_D}) and simulations of the full model agree well
close to the Hopf bifurcation, see Fig.~\ref{fig_motor_model}.
This minimal motor model 
recapitulates the experimental observation of phase-diffusion in a minimal setting
and illustrates how noisy oscillations can arise from small-number-fluctuations.

\textit{Flagellar synchronization.} Phase fluctuations cause phase-slips in pairs of synchronized flagella,
\textit{e.g.}\ in the green algae \textit{Chlamydomonas} \cite{Goldstein:2009}.
\textit{Chlamydomonas} swims with two flagella, which can synchronize their beat.
Analysis of phase-slips allowed a previous, indirect estimate
of flagellar phase fluctuations,
corresponding to $Q{\approx}25$ for the quality factor of individual flagella \cite{Goldstein:2009}.
A latter study indicated a length-dependence of $Q$,
with corresponding $Q$ ranging from ${\approx}70{-}120$ 
for length increasing from $6\micron$ to $12\micron$ \cite{Goldstein:2011}.
Interestingly, flagellar synchronization in \textit{Chlamydomonas} 
seems to operate just below a tolerable level of noise:
Consider the approximate dynamics of the phase difference $\delta$ 
between two identical, coupled oscillators, 
$\dot{\delta}=-\lambda/T\sin\delta+\zeta$, 
where $\zeta$ is Gaussian white noise with 
$\langle\zeta(t)\zeta(t')\rangle=4D\delta(t{-}t')$ \cite{Stratonovich:1963,Goldstein:2009}.
Using the estimate $\lambda{\approx}0.3$
for the synchronization strength \cite{Goldstein:2009},
we find $\lambda Q{\approx}10$, which yields robust synchronization. 
A ten-fold higher noise level, however, implies failure of synchronization, see Fig.~\ref{fig_path}B.

\textit{Conclusion.}
The beating flagellum is a noisy oscillator, 
driven by $N{\approx}8{\cdot}10^4$ dynein motor domains \cite{Nicastro:2006}.
Here, we precisely measured its phase and amplitude fluctuations,
using a novel method of limit-cycle reconstruction \cite{Geyer:2013}. 
We obtain a quality factor $Q{=}38{\pm}16.7$ of flagellar oscillations.
Values estimated in other cytoskeletal oscillators are
$Q{=}2.2{\pm}1.0$ ($N{\approx}2500$) for spontaneous hair bundle oscillations \cite{Barral:2010}, and
$Q{=}1.4{\pm}1.1$ ($N{=}10{-}100$) for an \textit{in-vitro} acto-myosin system \cite{Placais:2009}.
We find that the strength of flagellar phase fluctuations is several orders-of-magnitudes above 
the level corresponding to thermal noise, highlighting the active origin of flagellar fluctuations.

We compute the quality factor $Q$ in a minimal model of motor-filament oscillations,
and find that $Q$ is proportional to the number of motors.
A simple numerical example 
\endnote{
$N{=}8{\cdot}10^4$,
$l{=}8\mathrm{nm}$, 
$k{\approx}2\mathrm{nN}/\micron$ \cite{Howard:2001},
$\xi{\approx}40\mathrm{pN}\mathrm{s}/\micron$,
$\Omega{=}10^4\mathrm{s}^{-1}$,
$\xi_a/\xi{=}2$, 
$\alpha{=}\eta{=}0.5$.
}
yields noisy oscillations with 
amplitude, frequency, and quality factor,
$Al{\approx} 68\,\mathrm{nm}$,
$\omega_0{\approx} 228\,\mathrm{s}^{-1}$,
$Q{\approx} 33$, 
which roughly match measured values
($Al{\approx}100\,\mathrm{nm}$,
$\omega_0{\approx}200\,\mathrm{s}^{-1}$ \cite{Riedel:2007},
$Q{\approx}38$).
Our analytic approximation eq.~(\ref{eq_D}) is not applicable 
for these large-amplitude-oscillations.
Note that the model does not fully capture flagellar oscillations 
quantitatively because it strongly simplifies flagellar geometry and motor dynamics.

We show that phase and amplitude fluctuations affect sperm swimming differently:
Whereas amplitude fluctuations cause an effective diffusion of sperm swimming circles,
phase fluctuations imply speed fluctuations, but do not change the shape of the path.
Additionally, phase fluctuations introduce phase-slips in collections of synchronized flagella \cite{Goldstein:2009}.

\textit{Acknowledgments.}
We thank J Baumgart, VF Geyer, J Howard, P Romanczuk, P Sartori, T Schwalger for stimulating discussions.
Experimental data was recorded previously by IH Riedel-Kruse in the laboratory of J Howard.

Our analysis coarse-grains a phase-dependence of flagellar noise, but see \cite{Wan:2014}.

\def\urlprefix{}
\def\url#1{}
\bibliography{../../../bibliography/library}

\newpage

\onecolumngrid
\appendix

\renewcommand{\thesection}{\arabic{section}}
\renewcommand{\thesubsection}{\thesection.\arabic{subsection}}
\renewcommand{\thesubsubsection}{\thesubsection.\arabic{subsubsection}}

\makeatletter
\renewcommand{\p@subsection}{}
\renewcommand{\p@subsubsection}{}
\makeatother

\renewcommand{\theequation}{S\arabic{equation}}
\setcounter{equation}{0}

\begin{center}
\large
Appendix
\end{center}

We present details on the determination of the quality factor of noisy motor-filament oscillations
for the minimal motor model discussed in the main text.
We first derive a Langevin equation [see Eq.~(\ref{eq:Langevin_final})] for the stochastic motor dynamics
using methods presented in \cite{Guerin:2011a,Guerin:2011b}.
We then show how this Langevin equation can be transformed into 
a stochastic Hopf normal form using a center manifold technique,
see Eq.~(\ref{eq:Hopf_normal_Z_stoch}).
From this, we obtain an approximation for the quality factor,
see Eq.~(\ref{eq:quality_factor}).

In the minimal motor model \cite{Julicher:1997,Guerin:2011a}, 
a collection of $N$ motors is rigidly connected to a common backbone at equally spaced  positions $x_i=il/N$,
see Fig.~2A.
These motors interact with a filament of periodicity $l$:
Individual motors can bind and unbind from the filament with position-dependent transition rates
\begin{align}
\omegaon(x) &=\Omega[\eta-\alpha \cos(2\pi x/l)], \\
\omegaoff(x) &=\Omega-\omegaon.
\end{align}
Here, $\Omega$ denotes a characteristic transition rate, 
$\eta$ the duty ratio of motors, 
and $x$ a coordinate along the filament,
while $\alpha$ characterizes spatial variation of the transition rates. 
Note that $\omegaon(x)+\omegaoff(x)=\Omega$. 
This so-called uniform rate assumption greatly 
simplifies the analytical treatment of the model~\cite{Guerin:2011a}. 
Motors bound to the filament are subject to an interaction potential 
\begin{equation}
W(x)=U[1-\cos(2\pi x/l)].
\end{equation}
The filament is connected to the motor backbone via an elastic spring of stiffness $k = Nk_0$
and a dashpot with drag coefficient $\xi=N\xi_0$ operating in parallel, see Fig.~2A.
The dynamics of the filament is now given by 
\begin{equation}
kX+\xi\dot{X}=-\sum_{i} \partial_X W(x_i-X).
\end{equation}
To properly define a thermodynamic limit for large $N$,
we will rescale stiffness and viscosity as $k=k_0 N$ and $\xi=\xi_0 N$.

\section{Fokker-Planck equation of motor-filament dynamics}

We now derive a continuum description for the dynamics of the discrete set of motors.
In order to define a probability density $\rho_0(z)$ of bound motors, 
we divide the interval $[0,l]$ into $m$ bins of width $\Delta=l/m$ and respective bin centers $z_i=i\Delta-\Delta/2$,
and set 
$\rho_0(z_i)=(1/\Delta)n_i/N$,
where $n_i$ denotes the number of bound motors within the $i$-th bin.

Following \cite{Guerin:2011b}, 
we can formulate a master equation that governs the evolution of the joint probability distribution 
$P(X,n_1,\ldots,n_m)$ 
for the filament position $X$ and the bin counts $n_i$
\begin{align}
\label{eq:master}
\frac{\partial P}{\partial t} 
= & -\frac{\partial }{\partial X}
\left[\left(\sum_{i=1}^m \frac{W'(z_i-X)}{\xi_0}\frac{n_i}{N}-\frac{k_0}{\xi_0}X
\right)P\right ] \nonumber \\
& +\sum_{i=1}^m \omegaoff(z_i-X)({\bf E}_i^+ -1)\,n_i P \\
& +\sum_{i=1}^m \omegaon (z_i-X)({\bf E}_i^- -1)\left(\frac{N}{m}-n_i\right)P.
\nonumber
\end{align}
Here, ${\bf E}_i^{\pm}$ denote step operators, 
whose action on any function $f(n_i)$ obeys ${\bf E}_i^{\pm}f(n_i)=f(n_i\pm1)$.
Using bin center positions as approximate motor positions introduces a relative coarse-graining error $\mathcal{O}(1/m^2)$. 
To obtain a Fokker-Planck equation for $\rho_0(z)$, 
we replace $P(X,\{n_i\})$ by $P(X,\rho_0)$, 
expand Eq.~(\ref{eq:master}) using the operator expansion
\begin{equation}
\label{eq:Expansion}
{\bf E}_i^{\pm}=1\pm\frac{\partial}{\partial n_i}
+\frac{1}{2}\frac{\partial^2}{\partial n_i^2}\pm\cdots
\end{equation} 
and neglect all derivatives higher than the second order,
which implies a truncation error of order 
$\mathcal{O}(1/N^2)$, 
as well as a coarse-graining error of relative order $\mathcal{O}(1/m^2)$.
For further simplification, 
we change the reference frame from the common motor backbone to the co-moving frame of the filament,
and use henceforth the density $\rho(x)$ of bound motors with respect to the filament coordinate,
$\rho(x)=\rho_0(z-X)$
(where $\rho(x)$ shall be extended outside the interval $[-X,l-X]$ by periodic continuation for mathematical convenience).
This finally leads to a functional Fokker-Planck equation 
for the distribution function $P(X,\rho)$ (see also \cite{Guerin:2011b})
\begin{eqnarray}
\label{eq:FFP}
\frac{\partial P}{\partial t} & = &
-\frac{\partial }{\partial X} v P 
+\int_0^l \! dx\,
\frac{\delta}{\delta \rho(x)} A P  \\
&+&\frac{1}{2N}\int_0^l\!dx\, \int_0^l\! dy\,
\delta(x-y)\frac{\delta^2}{\delta \rho(x)\delta \rho(y)} C P.
\nonumber 
\end{eqnarray}
The drift terms read
\begin{align}
v&=\int_0^l \!dx\, \frac{W'(x)}{\xi_0}\rho(x)-\frac{k_0}{
\xi_0}X, \label{eq:drift_1}\\
A&=\omegaoff(x)\rho(x)-\omegaon(x)[1/l-\rho(x)]-v\partial_x\rho(x),
\label{eq:drift_2}
\end{align}
while the diffusion term reads
\begin{equation}
C=\omegaoff(x)\rho(x)+\omegaon(x)[1/l-\rho(x)].
\end{equation}
Choosing a bin size $m\sim \sqrt{N}$ that increases with the number of motors, 
we find that both drift terms and diffusion terms in Eq.~(\ref{eq:FFP}) are 
each accurate to leading order in $1/N$. 

\subsection{Spatial Fourier Expansion}

We expand $\rho(x)$ into a spatial Fourier series
\begin{equation}
\label{eq:Fourier_Series}
\rho(x)=\frac{\eta}{l} a_0 +
\frac{\alpha}{l} \sum_{n=1}^{\infty}
 a_n \cos\left(2\pi n\frac{x}{l}\right)
+b_n \sin\left(2\pi n\frac{x}{l}\right),
\end{equation} 
and rewrite the functional Fokker-Planck equation (\ref{eq:FFP})
in terms of the Fourier coefficients \cite{Guerin:2011b}
\begin{align}
\label{eq:FP_full}
\frac{\partial}{\partial t}P(\{a_n, b_n\}, X, t) 
=&
-\frac{\partial }{\partial X}
(v P)-\sum_n\left (\frac{\partial}{\partial a_n} A_n P
+\frac{\partial}{\partial b_n} B_n P\right)\nonumber \\
&
+\sum_{m,n}
\frac{\partial^2 D_{mn}^{aa} P}{\partial a_m\partial a_n} 
+ 2\frac{\partial^2 D_{mn}^{ab}P}{\partial a_m\partial b_n}
+\frac{\partial^2 D_{mn}^{bb}P}{\partial b_m\partial b_n}.
\end{align}
The drift terms $A_n$, $B_n$ characterize the deterministic mean-field dynamics of the system, and read
\begin{equation}
\label{eq:drift_terms}
A_n=\frac{2}{\alpha}\int_{0}^l A(x)\cos(2\pi n x/l) dx,\quad
B_n=\frac{2}{\alpha}\int_{0}^l A(x)\sin(2\pi n x/l) dx
\end{equation}
for $n\ge 1$, while $A_0=(1/\eta)\int_0^l A(x) dx$ and $B_0=0$.
The elements of the diffusion matrix characterize the noise effect due to a finite number of motors, and read
\begin{align}
\label{eq:diffusion_terms}
D_{mn}^{aa}&=\frac{2}{N\alpha^2} \int_{0}^l C(x)\cos(2\pi m x/l)
\cos(2\pi n x/l) dx, \nonumber \\
D_{mn}^{ab}&=\frac{2}{N\alpha^2} \int_{0}^l C(x)\cos(2\pi m x/l)
\sin(2\pi n x/l) dx, \\
D_{mn}^{bb}&=\frac{2}{N\alpha^2} \int_{0}^l C(x)\sin(2\pi m x/l)
\sin(2\pi n x/l) dx. \nonumber
\end{align}
In general, the noise strengths are state-dependent.
For small oscillation amplitudes and in the limit of weak noise,
we can approximate them by their respective values evaluated at the fixed point of the deterministic dynamics,
characterized by $a_0=1$, $a_1=-1$, $b_1=0$. 
Specifically, we find
\begin{align}
\label{eq:diffusion_constants_2}
D_a
=D^{aa}_{11}
&=\frac{\Omega}{2N}
\left[ 
\frac{2\eta}{\alpha^2}(1+a_0(1-2\eta))+3a_1
\right]
\approx
\frac{\Omega}{2N}\left[\frac{4\eta(1-\eta)}{\alpha^2}-3 \right],\nonumber\\
D_b
=D^{bb}_{11}
&=\frac{\Omega}{2N}
\left[
\frac{2\eta}{\alpha^2}(1+a_0(1-2\eta))+a_1 
\right]
\approx\frac{\Omega}{2N}\left[\frac{4\eta(1-\eta)}{\alpha^2}-1 \right],\\
D^{ab}_{11}
&=\frac{\Omega}{2N}b_1\approx 0.\nonumber
\end{align}

Remarkably, the dynamics of the principlal Fourier modes $a=a_1$, $b=b_1$, 
and filament position $X$ decouples from the other modes \cite{Guerin:2011b}
with corresponding Langevin dynamics
\begin{align}
&\dot{a}=-\Omega (a +1- \gamma b^2+\beta bX/l)+\zeta_a(t), \nonumber \\
\label{eq:Langevin_final}
&\dot{b}=-\Omega (b+\gamma b a-\beta a X/l)+\zeta_b(t), \\
& \dot{X}=\frac{\Omega l}{2\pi}(\gamma b-\beta X/l), \nonumber
\end{align}  
where $\zeta_i(t)$ denote Gaussian white noise terms satisfying 
$\langle \zeta_i(t)\zeta_j(t) \rangle=2D_{i}\,\delta_{ij}\delta(t-t')$ 
for $i,j=a,b$, and
$\beta/(2\pi)=\nu=k_0/(\xi_0\Omega)$,
$\gamma=\xi_a/\xi_0=2\pi^2\alpha U/(\Omega l^2 \xi_0)$.

We now show how Eq.~(\ref{eq:Langevin_final}) can be transformed into Hopf normal form.
We first treat the noise-free case, $D_a=D_b=0$.
We first do a linear transformation of the coordinate tuple $(a, b, X)$ 
to a new set of coordinates, comprising a real variable $y$ and a complex variable $Y$, 
\begin{equation}
\label{eq:linear_transformation}
\begin{pmatrix}
a+1\\2b\\2X/l
\end{pmatrix}=
\begin{pmatrix}
1 & 0 &0\\
0 & \chi & \chi^*\\
0 & 1 & 1
\end{pmatrix}
\begin{pmatrix}
y\\Y^*\\Y
\end{pmatrix},
\end{equation}
where
$\chi=\pi(-\epsilon+2\nu+2i\sqrt{\nu})/(\sqrt{\nu}+i)^2$, 
$\epsilon=\gamma-1-\nu$.
Conversely,
$Y=i(b-\chi X/l)/\mathrm{Im}\chi$
with 
$\mathrm{Im}\chi=-2\pi/(\sqrt{\nu}+1/\sqrt{\nu})+\mathcal{O}(\epsilon)$.
In the new coordinate set, 
the linearized dynamics at the fixed point 
$(y,Y)=(0, 0)$ 
is diagonal
\begin{align}
\label{eq:linear}
\frac{d}{dt}\begin{pmatrix}
y\\Y
\end{pmatrix}=
\begin{pmatrix}
-\Omega & 0\\
0 & \Omega(\epsilon/2+i\sqrt{\nu})
\end{pmatrix}
\begin{pmatrix}
y\\Y
\end{pmatrix}
\end{align}
One can show that $y$ relaxes to an invariant manifold $y=y(Y,Y^*)$ 
that is tangential to the plane $y=0$ at $(y,Y)=(0,0)$.
For this so-called center manifold~\cite{Crawford:1991}, we make a quadratic ansatz 
\begin{equation}
\label{eq:center_manifold}
y=h_1 Y^2 + h_1^* Y^{*2} + h_2 YY^* + \mathcal{O}(|Y|^3)
\end{equation}
with complex coefficients $h_i$ that can be determined self-consistently 
from the full nonlinear dynamics. 
The dynamics of $Y$ on the manifold defined by (\ref{eq:center_manifold}) 
comprises a linear term, as well as cubic terms as leading order nonlinearity
\begin{align}
\label{eq:Hopf_1}
\frac{dY}{dt}=
\Omega\left(\frac{\epsilon}{2}+i\sqrt{\nu}\right)Y
-g_0 Y^3 -g_1 Y^2 Y^* -g_2 Y Y^{*2} -g_3 Y^{*3}+\mathcal{O}(|Y|^4), \nonumber \\
\end{align}
where $g_i$ are complex numbers. 
Using nonlinear coordinate 
transformations of the form
$Y=Z+\theta Z^pZ^{*(3-p)}$,
all cubic nonlinearities can eliminated, with the exception of $Z^2Z^*$.
Thus, we have brought the dynamics of $Z$ into Hopf normal form 
\begin{align}
\label{eq:Hopf_normal_Z}
\frac{dZ}{dt}=\mu(\Lambda-|Z|^2)Z+i(\omega_c-\omega_1|Z|^2)Z+\mathcal{O}(|Z|^4),
\end{align}
with parameters
\begin{align}
\mu=\frac{3\pi^2\Omega\nu(1+2\nu)}{2(1+4\nu)},\quad 
\Lambda=\frac{\Omega\epsilon}{2\mu}=
\frac{1}{\pi^2}\frac{\epsilon(1+4\nu)}{3\nu(1+2\nu)},
\quad \omega_c&=\Omega\sqrt{\nu},\quad \omega_1=-\frac{\mu \sqrt{\nu}}{1+2\nu}.
\end{align}
For $\epsilon>0$, 
in the absence of noise, 
the complex oscillator variable 
$Z=A\exp i\varphi$ oscillates with amplitude
$A=|Z|=\sqrt{\Lambda}$ 
and frequency
$\omega_0=\omega_c-\omega_1\Lambda$.

In the case of weak noise, 
we can apply the same series of coordinate transformations used above
to the Langevin equation (\ref{eq:Langevin_final}),
while neglecting noise-induced drift terms of order $\mathcal{O}(1/N)$
\begin{align}
\label{eq:Hopf_normal_Z_stoch}
\frac{dZ}{dt}=\mu(\Lambda-|Z|^2)Z+i(\omega_c-\omega_1|Z|^2)Z+i\zeta(t),
\end{align}
where $\zeta(t)$ denotes Gaussian white noise with
$\langle\zeta(t)\zeta(t')\rangle=4D_0\Lambda\delta(t-t')$ 
and noise strength
\begin{equation}
\label{eq:noise}
4D_0\Lambda=2D_b\left(\frac{\sqrt{\nu}+1/\sqrt{\nu}}{2\pi}\right)^2+\O(\epsilon).
\end{equation}

We now compute the variance of amplitude fluctuations and the phase diffusion coefficient.
We consider the limit of weakly perturbed oscillations,
$\sigma_A^2\ll\Lambda$.
Using Stratonovich calculus, 
we derive from Eq.~(\ref{eq:Hopf_normal_Z_stoch}) 
equations for the instantaneous amplitude $A$ and phase $\phi$
\begin{align}
\label{eq:Langevin_A_phi}
\dot{A} &=\mu(\Lambda-A^2)A+\sin\phi\,\zeta(t),\\
\dot{\phi}&=\omega_c-\omega_1 A^2+\frac{\cos\phi}{A}\zeta(t).
\end{align}
We approximate the phase-dependent noise strengths by
their phase-averaged expectation values,
which will reproduce, to leading order in the noise-strength, 
the same dynamics on time-scales longer than the oscillation period.
We also linearize the stochastic dynamics Eq.~(\ref{eq:Langevin_A_phi}) for small amplitude fluctuation $\dA$, 
neglecting terms of order $\mathcal{O}(\dA^2)$,
\begin{align}
\label{eq:Langevin_A_phi_lin}
\frac{d}{dt}\dA &\approx-2\mu\Lambda\dA+\frac{1}{\sqrt{2}}\zeta(t),\\
\dot{\phi}&\approx \omega_c-\omega_1\Lambda-2\omega_1\sqrt{\Lambda}\,\dA+\frac{1}{\sqrt{2\Lambda}}\zeta(t).
\end{align}
The first equation describes an Ornstein-Uhlenbeck process 
with correlation time $\tau_A=(2\mu\Lambda)^{-1}$
and variance 
\begin{equation}
\sigma_A^2=D_0\Lambda\tau_A=D_0/(2\mu).
\end{equation}

For the phase-diffusion coefficient, we find
\begin{align}
D 
&=\lim_{t\rightarrow\infty} \frac{1}{2t} 
\left( \langle[\varphi(t)-\varphi(0)]^2\rangle-\langle[\varphi(t)-\varphi(0)]\rangle^2 \right) \\
&=\lim_{t\rightarrow\infty} \frac{1}{2t} 
\langle\int_0^t\!\!\int_0^t\! dt_1 dt_2 \, \dot{\varphi}(t_1)\dot{\varphi}(t_2)\rangle-\omega_0^2 \\
&= \left[ 1 + \left(\frac{\omega_1}{\mu}\right)^2\right] D_0.
\end{align}

We now readily find for the qualify factor 
\begin{eqnarray}
\label{eq:quality_factor}
Q=\frac{\omega_0}{2D}=\Theta\frac{\omega_0}{2\Omega}\frac{N\Lambda}{\eta(1-\eta)}
\left(\frac{2\pi\alpha}{\sqrt{\nu}+1/\sqrt{\nu}}\right)^2
\end{eqnarray}
with prefactor 
\begin{equation}
\Theta=
\left[1-\frac{\alpha^2}{4\eta(1-\eta)}\right]\left[1+\left(\frac{\omega_1}{\mu}\right)^2\right].
\end{equation}
This prefactor can be shown to vary around 1 within close bounds, 
\begin{equation}
\label{eq:Theta_ineq}
3/4\leq \Theta \leq 9/8,
\end{equation}
and has therefore been omitted in the approximation presented in the main text.
The proof of inequality~(\ref{eq:Theta_ineq}) involves 
$0\leq\alpha \leq\eta$ and $\alpha\leq(1-\eta)$, 
as well as  
$|\omega_1/\mu|=\sqrt{\nu}/(1+2\nu)=2^{-1/2}/[(2\nu)^{-1/2}+(2\nu)^{1/2})\leq 2^{-3/2}$.

This approximation is only valid for weakly perturbed oscillations with $\sigma_A^2\ll\Lambda$;
the latter condition can be rephrased as $N\gg 1/\epsilon^2$.
We remark that amplitude fluctuations remain finite, even at the Hopf bifurcation,
and can be shown to scale as $\sigma_A^2\sim N^{-1/2}$ for $\epsilon=0$.

\end{document}